\documentclass[aps,prd,groupedaddress,graphicx,nofootinbib]{revtex4-2}
\usepackage{amsmath,amssymb,graphics,graphicx,color,epsf}
\usepackage{subfigure}
\usepackage[scanall]{psfrag}  
\usepackage{tikz}

\begin{document}

\title{More solutions for the Wheeler-DeWitt equation in a flat FLRW minisuperspace}

\author{Chia-Min Lin}

\affiliation{Fundamental General Education Center, National Chin-Yi University of Technology, Taichung 41170, Taiwan}



\begin{abstract}
This work proposes more solutions for the Wheeler-DeWitt equation in a flat FLRW minisuperspace. We study quantum cosmology in the framework of the de Broglie-Bohm interpretation and investigate the quantum cosmological effects throughout the evolution of the universe. In a particular solution, the tendency for a scalar field to roll down the potential is balanced by the quantum force, and a Minkowski spacetime is obtained.

\end{abstract}
\maketitle
\large
\baselineskip 18pt
\section{Introduction}
\label{sec1}

Quantum cosmology (see \cite{Kolb:1990vq, Linde:1990flp, Halliwell:1989myn, Wiltshire:1995vk} for some reviews and more references therein) is an ambitious subject. The idea is to apply quantum mechanics to the universe as a whole. Usually, quantum mechanics is considered on a microscopic scale. On the other hand, classical physics is applied on a macroscopic scale. By definition, the most macroscopic thing in physics is the whole universe. Is it really make sense to apply quantum mechanics to the whole universe?

Through the idea of cosmic inflation, our universe, although incredibly huge now, was exponentially small in the past. Indeed it is possible to emerge from the spacetime foam with size compared to the Planck length. At least quantum cosmology seems relevant at the creation of the universe or when the energy density is close to the Planck scale.

In quantum mechanics, there is a Schr\"{o}dinger equation to describe the wave function $\Psi$ of a particle. The usual interpretation is that $|\Psi|^2$ gives the probability of finding the particle in a particular place. In quantum cosmology, as illustrated in more detail in Section \ref{wdes}, there is a Wheeler-DeWitt equation to describe the wave function $\Psi$ of a universe. However, there is currently no consensus about the meaning of $|\Psi|$ (for example, see \cite{Shestakova:2019rsh} for a discussion). Can we interpret $|\Psi|$ as the probability of the creation of a universe? But $\Psi$ does not depend on time; how could it possibly know when the time of creation is? The study of quantum cosmology often brings more questions than that being answered, but this is why the subject is interesting. 

My strategy is to come up with simple and solvable models. We can clearly see how the evolution of the universe described by Einstein gravity is modified by the quantum cosmological effects. We choose to work on quantum cosmology in the framework of de Broglie-Bohm interpretation of quantum mechanics. One reason is that it avoids problems such as how wave function collapsing happens. However, the solution we found is valid for the Wheeler-DeWitt equation, and they are subject to other interpretations according to readers' tastes.
Our model may be helpful in addressing questions such as whether quantum cosmology is relevant during the evolution of the universe. Or is there any quantum cosmology effect when the energy density is much lower than the Planck scale? In this work, we try to show that the answer seems affirmative.

\section{The Wheeler DeWitt equation}
\label{wdes}
The Wheeler-DeWitt equation in its original form is a complicated functional differential equation for the wave functional $\Psi$ defined on a space known as a superspace\footnote{Superspace here has nothing to do with supersymmetry.}. In practice, the infinite degrees of freedom of the superspace are restricted into a finite-dimensional subspace called minisuperspace, such as a homogeneous and isotropic universe.

We consider a minisuperspace, which is a flat Friedmann-Lema\^{i}tre-Robertson-Walker (FLRW) universe with the metric
\begin{equation}
ds^2=N^2dt^2-a^2d\Omega^2_k.
\end{equation}
Here $N$ is the lapse function, $a\equiv e^\alpha$ is the scale factor, and $d\Omega^2_k$ is the spatial line element with curvature $k=-1,0,+1$. Note that $\dot{\alpha}$ is the Hubble parameter. In the following, we will choose the gauge $N=1$ (this makes $t$ so-called cosmic proper time). 
It is typical to consider a close FLRW universe in the context of quantum cosmology (see \cite{Kolb:1990vq}). The reason is to apply quantum cosmology to the creation of a universe. On the other hand, we are interested in the quantum cosmological effects during the evolution of the universe. Therefore, we prefer to consider a flat universe\footnote{It is conceivable that a flat or open universe with non-trivial topology could emerge from nothing \cite{Coule:1999wg, Linde:2004nz}.}.
In fact, no experimental evidence indicates that our universe has ever been closed or open.
In the following discussion, we set the reduced Planck mass to unity, namely $1/8\pi G \equiv M_P^2=1$.
We consider a scalar field $\phi$ with potential $V$ as the matter field.
If there is a cosmological constant, we combine it into the constant part of $V$. Since $\sqrt{-g}=e^{3\alpha}$, the classical Lagrangian is
\begin{equation}
L=e^{3\alpha}\left( \frac{\dot{\phi}^2}{2}-3\dot{\alpha}^2-V \right).
\end{equation}
The momenta conjugate to $\alpha$ and $\phi$ are 
\begin{equation}
\pi_\alpha=\frac{\partial L}{\partial \dot{\alpha}}=-6e^{3\alpha}\dot{\alpha},   \;\;\;   \pi_\phi=\frac{\partial L}{\partial \dot{\phi}}=e^{3\alpha}\dot{\phi}.
\end{equation}
The corresponding Hamiltonian is
\begin{equation}
H=e^{-3\alpha}\left( \frac{1}{2}\pi_\phi^2-\frac{1}{12}\pi_\alpha^2+V \right).
\end{equation}
After canonical quantization, $\pi_\alpha \rightarrow -i\partial/\partial \alpha$ and $\pi_\phi \rightarrow -i\partial/\partial \phi$ the Wheeler-DeWitt equation, $H\Psi(\alpha, \phi)=0$ for the wave function of the universe is given by \cite{Lin:2023sza}
\begin{equation}
-\frac{1}{12}\frac{\partial^2}{\partial \alpha^2}\Psi+\frac{1}{2}\frac{\partial^2}{\partial \phi^2}\Psi +q \frac{\partial}{\partial \alpha}\Psi-e^{6\alpha}V \Psi=0.
\label{wde}
\end{equation}
Here the parameter $q$ is any real number that parameterizes the operator ordering ambiguity in the Hamiltonian for quantization. Some argue that we should have $q=0$ at the expense of introducing fermion degrees of freedom \cite{Hawking:1985bk}. However, the proposal appears to be not conclusive \cite{Vilenkin:1986cy}. In this work, we shall keep $q$ unspecified\footnote{In quantum cosmology, semi-classical calculations do not depend on the value of $q$ \cite{Kolb:1990vq}.}. In fact, we would eventually choose $q=1/4$. I want to find solutions for Eq.~(\ref{wde}) for a reasonable $V$.

\section{The de Broglie-Bohm interpretation of quantum mechanics}
\label{Bohm}
In this section, I briefly review the de Broglie-Bohm interpretation of quantum mechanics \cite{Bohm:1951xw, Bohm:1951xx} (see \cite{Bohm:1987np} for a review).
Let us consider the Schr\"{o}dinger equation for some particle with mass $m$
\begin{equation}
i \frac{\partial \Psi}{\partial t}=-\frac{\nabla^2 \Psi}{2m}+V\Psi,
\end{equation}
where I have set $\hbar=1$.
Substituting $\Psi=|\Psi|e^{iS}$ into the above equation, one can find
\begin{equation}
\frac{\partial S}{\partial t}=-\left[ \frac{(\nabla S)^2}{2m}+V-\frac{\nabla^2 |\Psi|}{2m |\Psi|} \right].
\end{equation}
This can be interpreted as a (generalized) Hamilton-Jacobi equation, where the velocity $v$ of the particle is 
\begin{equation}
v=\frac{\nabla S}{m}.
\label{v}
\end{equation}
This is known as the guidance equation. The trajectory $x(t)$ can be obtained by integrating the guidance equation. However, the initial condition is regarded as the hidden variable of the theory. 
A new potential called the quantum potential is defined as 
\begin{equation}
Q \equiv -\frac{\nabla^2 |\Psi|}{2m|\Psi|}.
\label{qup}
\end{equation}
This allows us to write the equation of motion as
\begin{equation}
m\frac{d^2x}{dt^2}=-\nabla V-\nabla Q,
\end{equation}
where $-\nabla Q$ is called the quantum force. As can be seen from Eq.~(\ref{qup}), even if $\Psi$ is small, the quantum potential can be very high. In some sense, $\Psi$ provides the ``information" to guide the movement of the particle, like a radio signal guides an automatic vehicle. The difference between classical and quantum physics is due to the quantum potential $Q$. If $Q=0$, the classical limit is obtained. Note whether or not we consider de Broglie-Bohm interpretation $\Psi$ is the solution of the Schr\"{o}dinger equation. This, in turn, determines $Q$.

There is also the continuity equation
\begin{equation}
\frac{\partial P}{\partial t}+ \nabla \cdot \left( P \frac{\nabla S}{m} \right)=0,
\label{continuity}
\end{equation}
where $P \equiv |\Psi|^2$. Here $P$ is not necessarily the probability density. If a statistical ensemble of particles have a probability density equal to $P$, it is said to be in a state of quantum equilibrium. However, a systerm is allowed to deviate from quantum equilibrium and this provides an experimental possibility to distinguish de Broglie-Bohm interpretation from others.

In this interpretation, there is no uncertainty in determining the position and momentum of the particle. Therefore the theory is deterministic and ontological, like classical physics. In this interpretation, the position and momentum of the particle is there whether there is an observer or not. The concept of beable replaces the concept of observable. How could it be possible? The uncertainty of quantum mechanics actually hides in determining the initial condition of the particle, which can be regarded as hidden variables. According to Bell's theorem \cite{Bell:1987hh}, such theory must be nonlocal. The de Broglie-Bohm theory is advertised by Bell in \cite{Bell:1982xg}.

Let us consider an example that is taken from \cite{Bohm:1951xw}. The ``s'' state wave function of a hydrogen atom is 
\begin{equation}
\Psi = f(r) e^{i(\beta-Et)},
\end{equation}
where $\beta$ is an arbitrary constant and $r$ is the radius. We have
\begin{equation}
S=\beta-Et.
\end{equation}
Therefore, from Eq.~(\ref{v}),
\begin{equation}
v=0.
\end{equation}
In this interpretation, the electron in this state is standing still. The quantum force balances the electric force.
We will see in Section \ref{more} an analogous example in quantum cosmology.

\section{de Broglie-Bohm quantum cosmology}

As discussed in the introduction section, there are conceptual problems in applying the usual probability interpretation of quantum mechanics to the wavefunction of the universe. For example, in quantum mechanics, the probability amplitude is calculated at a time, given a specific state at a previous time. However, such time is absent for the Wheeler-DeWitt equation. In addition, there is no observer outside the universe to make classical measurements, which collapses the wavefunction in accordance with the Copenhagen interpretation. Furthermore, the meaning of a spacetime singularity (such as a big bang or big crunch) is ambiguous because there is no actual metric.  

These problems may be resolved if one considers quantum cosmology in the framework of the de Broglie-Bohm theory. The main reason is that the evolution of the universe is then described in terms of classical trajectories, independently from measurements or external observers, at the expense of introducing non-locality. Moreover, time plays the same role as in classical general relativity.

Starting from the Wheeler-DeWitt equation Eq.~(\ref{wde}) and assuming $\Psi=|\Psi|e^{iS}$, we can discuss quantum cosmology in the framework of de Broglie-Bohm interpretation  \cite{Vink:1990fm} (see \cite{Pinto-Neto:2004szq, Pinto-Neto:2013toa, Goldstein:1999my} for reviews).
The role played by the position of a particle in the previous section is replaced by the scalar field value $\phi$ and the parameter $\alpha$. One is for the matter field and the other is for gravity or geometry. The guidance equations give their time evolutions
\begin{equation}
\dot{\phi}=\frac{\partial_\phi S}{e^{3\alpha}},
\label{pdot}
\end{equation} 
and
\begin{equation}
\dot{\alpha}=-\frac{\partial_\alpha S}{6e^{3\alpha}}.
\label{adot}
\end{equation}
Assuming $\Psi=|\Psi|e^{iS}$. The derivatives with respect to $\alpha$ and $\phi$ produce
\begin{eqnarray}
\frac{\partial\Psi}{\partial \alpha}&=&\partial_\alpha |\Psi|e^{iS}+|\Psi|e^{iS}i \partial_\alpha S    ,\\
\frac{\partial^2 \Psi}{\partial \alpha^2}&=&\partial^2_\alpha |\Psi|e^{iS}+2\partial_\alpha |\Psi|e^{iS}i\partial_\alpha S+|\Psi|e^{iS}(i\partial_\alpha S)^2+|\Psi|e^{iS}i\partial^2_\alpha S   ,\\
\frac{\partial^2 \Psi}{\partial \phi^2}&=&\partial^2_\phi |\Psi|e^{iS}+2\partial_\phi |\Psi|e^{iS}i\partial_\phi S+|\Psi|e^{iS}(i\partial_\phi S)^2+|\Psi|e^{iS}i\partial^2_\phi S.
\end{eqnarray}
Substituting these results into Eq.~(\ref{wde}) and divided by $\Psi$, the real part gives
\begin{equation}
\frac{1}{12}(\partial_\alpha S)^2-\frac{1}{2}(\partial_\phi S)^2-e^{6\alpha}(V+Q_M+Q_G)=0,
\label{s}
\end{equation}
where
\begin{equation}
Q_M \equiv -\frac{1}{2e^{6\alpha}}\frac{\partial^2_\phi |\Psi|}{|\Psi|}, \mbox{ and} \;\;\; Q_G \equiv \frac{1}{12e^{6\alpha}}\frac{\partial^2_\alpha |\Psi|}{|\Psi|}-q \frac{\partial _\alpha |\Psi|}{e^{6\alpha}|\Psi|}
\label{qpo}
\end{equation}
are the quantum potentials. Here $Q_M$ corresponds to the quantum potential of the matter field, and $Q_G$ corresponds to the quantum potential of gravity. They represent the corrections from the quantum cosmological effects.
By using Eqs.~(\ref{pdot}) and (\ref{adot}) to eliminate $S$, we obtain
\begin{equation}
3 \dot{\alpha}^2=\frac{\dot{\phi}^2}{2}+V+Q_M+Q_G.
\label{mfri}
\end{equation}
This equation is just the classical Friedmann equation with the last three terms as its ``quantum corrections".
Note that the quantum potentials only depend on the shape of $|\Psi|$ instead of the magnitude of it. They provide active information for the time evolution of the relevant variables. This is a characteristic feature of the de Broglie-Bohm theory.
The equation of motion of the scalar field $\phi$ is given by 
\begin{equation}
\ddot{\phi}+3H\dot{\phi}+\partial_\phi(V+Q_M+Q_G)=0.
\end{equation}
One of the benefits of considering the de Broglie-Bohm interpretation is that the meaning of the classical limit is very clear; it simply means the cases where $Q_M=Q_G=0$.

\section{a solution}
\label{asol}
In \cite{Lin:2023sza}, for the matter field, I consider a homogeneous scalar field $\phi$ with a potential 
\begin{equation}
V=\frac{3\lambda^2}{4}\phi^2-\frac{\lambda^2}{2},
\end{equation}
where $\lambda$ is a parameter. This potential contains a mass term with the mass square $m^2=3\lambda^2/2$ and a negative cosmological constant that is related to the mass.
I present an analytical solution to the Wheeler-DeWitt equation Eq.~(\ref{wde}) as
\begin{equation}
\Psi=e^{-i\lambda e^{3\alpha}\phi}.
\label{sol1}
\end{equation}
From the guidance equation Eq.~(\ref{pdot}), we have
\begin{equation}
\dot{\phi}=-\lambda.
\label{uni}
\end{equation}
This implies the scalar field is rolling down the potential at a constant speed given by $\lambda$. Therefore, 
\begin{equation}
\phi=-\lambda t+c,
\label{sop}
\end{equation}
where $c$ is an integration constant. On the other hand, from the guidance equation Eq.~(\ref{adot}), we obtain
\begin{equation}
\dot{\alpha}=\frac{\lambda \phi}{2}=-\frac{\lambda^2}{2}t+\frac{\lambda}{2}c.
\end{equation}
Therefore
\begin{equation}
\alpha=-\frac{\lambda^2}{4}t^2+\frac{\lambda}{2}ct+d,
\end{equation}
where $d$ is another integration constant that can be set to $d=0$ by imposing $\alpha(0)=0$.
The scale factor is given by
\begin{equation}
a=e^\alpha=e^{-\frac{\lambda^2}{4}\left( t-\frac{c}{\lambda}\right)^2+\frac{c^2}{4}}.
\end{equation}
This describes a universe where the scale factor is given by a Gaussian function.
This solution is classical because from Eqs.~(\ref{qpo}), it is apparent that we have $Q_M=Q_G=0$ since $|\Psi|=1$.
Based on the classical nature of the solution, an inflation model is built upon it and is called the uniform rate inflation \cite{Lin:2023xgs} due to Eq.~(\ref{uni}).

\section{more solutions}
\label{more}
In the solution given by Eq.~(\ref{sol1}), $\lambda$ is a parameter. Therefore, changing $\lambda$ to $-\lambda$ is still a solution. Actually, the solutions correspond to time reversal solutions for the equation of motion of $\phi$. If one describes the expansion of the universe, the other describes the contraction of the universe.
More generally, we can consider solutions of the form
\begin{equation}
\Psi=Ae^{-i\lambda e^{3\alpha}\phi}+Be^{i\lambda e^{3\alpha}\phi},
\label{phi}
\end{equation}
where $A$ and $B$ are complex coefficients. 
Just like in Section \ref{Bohm}, whether or not we consider de Broglie-Bohm interpretation $\Psi$ is the solution of the Wheeler-DeWitt equation. 
Using the freedom to multiply $\Psi$ with an arbitrary complex number, we can make $A$ a real number which is smaller or equal to one and parameterize $A$ and $B$ as \cite{Vink:1990fm}
\begin{equation}
A=\cos \theta,   \;\;\;  B=\sin \theta \left( \cos \beta + i \sin \beta \right),  
\end{equation}
where $\theta \in [0,\pi] $ and $\beta \in [0,2\pi)$. 
Namely, we constrain $A$ and $B$ on the surface of a two-sphere in the four-dimensional space spanned by $\operatorname{Re}A$, $\operatorname{Im}A$, $\operatorname{Re}B$, and $\operatorname{Im}B$.
Therefore we have
\begin{equation}
\Psi= \cos \theta \cos \left( \lambda e^{3\alpha} \phi \right) + \sin \theta \cos \left( \lambda e^{3\alpha} \phi +\beta \right) -i \cos \theta \sin \left( \lambda e^{3\alpha} \phi \right) +i \sin \theta \sin \left( \lambda e^{3\alpha} \phi +\beta \right).  
\end{equation}
This implies
\begin{equation}
|\Psi|^2=1+2\cos \theta \sin \theta \cos \left( 2\lambda e^{3\alpha} \phi+\beta \right). 
\end{equation}
If we write $\Psi=|\Psi|e^{iS}$, then
\begin{equation}
\tan S= \frac{\tan \theta \tan \left( \lambda e^{3\alpha} \phi +\beta \right)-\sin \left( \lambda e^{3\alpha} \phi \right) }{\cos \left( \lambda e^{3\alpha} \phi \right) +\tan \theta \cos \left( \lambda e^{3\alpha} \phi +\beta \right) },
\end{equation}
and 
\begin{equation}
S= \tan^{-1}\left(  \frac{\tan \theta \tan \left( \lambda e^{3\alpha} \phi +\beta \right)-\sin \left( \lambda e^{3\alpha} \phi \right) }{\cos \left( \lambda e^{3\alpha} \phi \right) +\tan \theta \cos \left( \lambda e^{3\alpha} \phi +\beta \right) } \right).
\end{equation}
By using Eqs.~(\ref{pdot}) and (\ref{adot}), time evolutions of $\phi$ and $\alpha$ can be found.

\begin{figure}[t]
  \centering
\includegraphics[width=0.6\textwidth]{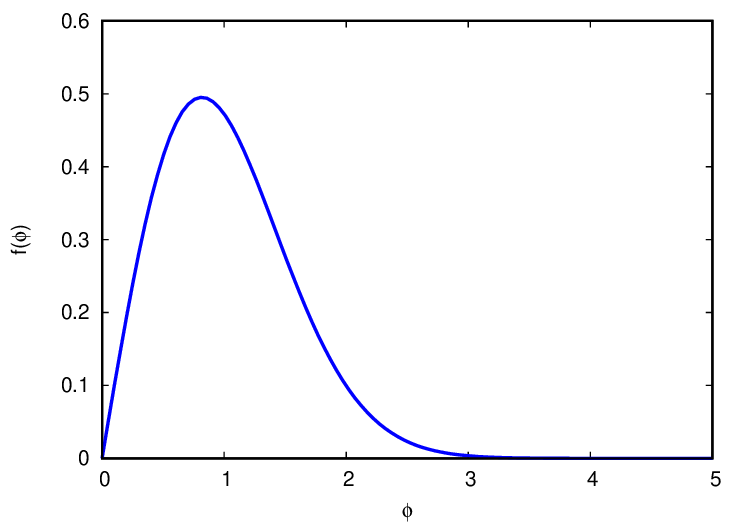}
  \caption{The function $f(\phi)$. We consider the region of $\phi$ where $f(\phi) \simeq 0$.}
  \label{fig1}
\end{figure}

In order to simplify the calculation, let us assume $\beta=0$, which means we consider the case where $A$ and $B$ are real numbers. This implies
\begin{equation}
\tan S= \tan \left( \lambda e^{3\alpha} \phi \right) \frac{\tan \theta-1}{\tan \theta+1} \equiv \tan \left( \lambda e^{3\alpha} \phi \right) \times \Theta,
\end{equation}
where we have defined a parameter $\Theta$, which can be any real number in the interval $(-\infty, +\infty)$.
We can write
\begin{equation}
S=\tan^{-1}\left( \Theta \tan \left( \lambda e^{3\alpha} \phi \right) \right).
\end{equation}
From Eq.~(\ref{pdot}), we have
\begin{equation}
\dot{\phi}=\frac{\partial_\phi S}{e^{3\alpha}}=\frac{\sec^2 \left( \lambda e^{3\alpha} \phi \right) \lambda \Theta}{\Theta^2\tan^2 \left( \lambda e^{3\alpha} \phi \right) +1},
\label{pd}
\end{equation}
and from Eq.~(\ref{adot}),
\begin{equation}
\dot{\alpha}=-\frac{\partial_\alpha S}{6 e^{3\alpha}}= -\frac{\partial_\phi S}{e^{3\alpha}}=-\frac{\sec^2  \left( \lambda e^{3\alpha} \phi \right) \lambda \Theta}{\Theta^2\tan^2 \left( \lambda e^{3\alpha} \phi \right) +1}\frac{\phi}{2}=-\frac{\dot{\phi}\phi}{2}=-\frac{1}{4}\frac{d}{dt}\left(\phi^2\right).
\label{ad}
\end{equation}
We can integrate to obtain
\begin{equation}
\alpha=-\frac{\phi^2}{4}+c,
\label{32}
\end{equation}
where $c$ is an integration constant.
By using the above result, Eq.~(\ref{pd}) can be written as
\begin{equation}
dt= \left[  \frac{\Theta}{\lambda}+\frac{1-\Theta^2}{\lambda \Theta}\cos^2 \left( \lambda \phi c e^{-\frac{3}{4}\phi^2}\right)   \right]d\phi.
\end{equation}
In order to further simplify the calculation, let us consider the region where the function $f(\phi)=\phi e^{-\frac{3}{4}\phi^2}$ can be approximated as zero. This function is plotted in Fig.~\ref{fig1}. In this region, we may set the cosine part to one (with reasonable $\lambda c$), and the integral can be done to obtain
\begin{equation}
\phi=\lambda \Theta t +c^\prime,
\label{msol}
\end{equation}
where $c^\prime=\lambda \Theta c$ is a constant. This can be compared with Eq.~(\ref{sop}).
We can see the role of $\Theta$ here in this region of solutions. It controls the rolling speed of the field $\phi$.
The solution in Section \ref{asol} corresponds to the case $A=1$ and $B=0$. This implies $\theta=0$ and $\Theta=-1$.
Note that $\Theta$ can be any real number. Therefore the constant rolling speed of $\phi$ can be adjusted to have any value as long as $\lambda$ is not zero. Each case has a corresponding Gaussian evolution of the scale factor $a=e^{\alpha}$ where $\alpha$ is given by Eq.~(\ref{32}).

One special case is $\Theta=0$. This means $A=B=\frac{1}{\sqrt{2}}$. From Eq.~(\ref{phi}), the wavefunction becomes
\begin{equation}
\Psi= \sqrt{2}\cos \left( \lambda e^{3\alpha} \phi \right),
\label{mins}
\end{equation}
which is real and $S=0$.
The field $\phi$ is frozen, as can be seen from Eq.~(\ref{pdot}) or Eq.~(\ref{msol}). In this case, from Eq.~(\ref{ad}), 
\begin{equation}
H=\dot{\alpha}=0.
\end{equation}
This can also be checked from Eq.~(\ref{mfri}) by calculating the corresponding $Q_M$ and $Q_G$.

We obtain a Minkowski spacetime. Classically the field $\phi$ should roll down the potential. However, there is a quantum force to stop it. The quantum force comes from the effect of quantum cosmology. This shows even a Minkowski spacetime could be nontrivial. To some extent, this is analogous to the hydrogen atom discussed in section \ref{Bohm}. In the ground state, the electron is static due to the balance between the electric and quantum forces. Here there is a balance between the force from the scalar potential and the quantum force from the quantum cosmological effect. One can imagine that since $A=B$, the expansion solution and the contraction solution has the same weight. Therefore, the universe neither expands nor contracts.

This solution shows that even in such Minkowski spacetime, quantum cosmological effects may play a role in the history of the universe.

\section{In comparison with the probability interpretation}

We focus on the de Broglie-Bohm interpretation of quantum cosmology in this work for the reasons explained in previous sections. However, our solutions are for the Wheeler-DeWitt equation. Therefore, they are valid for other interpretations. The probability interpretation faces problems, such as (the nonexistence of) external observers. We will ignore them in this section and try to contrast the differences between the interpretations. Defining the probability amplitude in quantum cosmology is a non-trivial task since the Wheeler-DeWitt equation is similar to the Klein-Gordon equation. Hence, there is a problem of negative probability. Here, we follow \cite{Vilenkin:1988yd}, where the probability of finding the universe in a surface element $d\Sigma_\alpha$ of a constant-``time" hypersurface $\Sigma_\alpha$ is defined as
\begin{equation}
dP=j^\alpha d\Sigma_\alpha,
\end{equation}
where the conserved current $j^\alpha$ is given by
\begin{equation}
j^\alpha=-\frac{i}{2}g^{\alpha\beta}\left( \Psi^\ast \nabla_\beta \Psi-\Psi \nabla_\beta \Psi^\ast \right).
\label{vp}
\end{equation}
The notation $\nabla_\alpha$ denotes a covariant derivative in the superspace metric\footnote{Sometimes it is called the DeWitt metric or supermetric. Here it is the metric of a minisuperspace.} which can be read off from Eq.~(\ref{s})
\begin{equation}
g^{\alpha \beta}=
\begin{bmatrix}
-\frac{1}{12}e^{-3\alpha} & 0 \\
0 & \frac{1}{2}e^{-3\alpha}.
\end{bmatrix}
\end{equation}
The conservation of the current
\begin{equation}
\nabla_\alpha j^\alpha=0 
\end{equation}
is the relativistic version of Eq.~(\ref{continuity}). If the hypersurface is chosen judiciously, the probability can be positive definite.
In our model, from Eq.~(\ref{phi}), we have
\begin{equation}
j^\phi=\frac{B^2-A^2}{2}\lambda,
\end{equation}
and
\begin{equation}
j^\alpha=\frac{A^2-B^2}{4}\lambda \phi.
\end{equation}
The hypersurface can be chosen to be the surfaces of constant $S$.
For example, for the solution of Minkowski spacetime with $A=B$, we have $j^\phi=j^\alpha=0$ and $P=0$. This can also be seen directly from Eq.~(\ref{vp}) since Eq.~(\ref{mins}) is real. This result might be plausible since if the range of $\alpha$ and $\phi$ are infinity, the probability of finding a particular (or some finite range) of $\alpha$ and $\phi$ should be (or approaches to) zero. On the other hand, in de Broglie-Bohm interpreration, a state is not an observable; but a beable. Probabilities are not fundamental in this theory; we only have a single universe. 

\section{conclusion}
This work considers the Wheeler-DeWitt equation in a flat FLRW minisuperspace. A solution was found in my previous work \cite{Lin:2023sza}. Here more solutions of the Wheeler-DeWitt equation are found. The new solutions are more versatile and contain different behaviors of the scale factor. In some solution regions, the scalar field's rolling rate is a constant with a speed that can take almost any value with a non-zero $\lambda$. A particular solution describes the scalar field $\phi$ rolling at zero rates, standing still like the electron in the ground state of the hydrogen atom. This universe is neither expanding nor contracting. It is just a Minkowski spacetime. Therefore even Minkowski spacetime may be a nontrivial solution. There is a balance between the gradient of the potential $V$ of $\phi$ and a quantum force.

These solutions explicitly show that quantum cosmological effects may be relevant throughout the evolution of the universe, even if the energy density (of the matter field) is small compared with the Planck scale. This work suggests that quantum cosmology is not only about the creation of the universe; it may play an important role in describing a macroscopic universe.

\section{data availability statement}
The authors confirm that the data supporting the findings of this study are available within the article.
\acknowledgments
This work is supported by the National Science and Technology Council (NSTC) of Taiwan under Grant No. NSTC 111-2112-M-167-002.

\end{document}